\documentclass[10pt]{iopart}
\usepackage{graphicx}
\usepackage{iopams}
\usepackage{setstack}
\usepackage[active]{srcltx}

\begin{document}

\title{The dynamic conductivity of strongly non-ideal plasmas: is the Drude
model valid?}
\author{ V.~M.~Adamyan$^{1}$, A.~A.~Mihajlov$^{2}$, N.~M.~Sakan$^{2}$,
        V.~A.~Sre\'{c}kovi\'{c}$^{2}$ and I.~M.~Tkachenko$^{3}$}
       {\address{$^{1}$Department of Theoretical Physics, Odessa National
        University, 65026 Odessa, Ukraine}
       \address{$^{2}$Institute of Physics, Pregrevica 118, Zemun,
        Belgrade, Serbia} }
       \address{$^{3}$Instituto Universitario de Matem\'{a}tica Pura y Aplicada,
         Universidad Polit\'{e}cnica de Valencia, 46022 Valencia, Spain}
\ead{vlada@ipb.ac.rs}

\begin{abstract}
The method of moments is used to calculate the dynamic conductivity
of strongly coupled fully ionized hydrogen plasmas. The electron
density $n_{e}$ and temperature $T$ vary in the domains $ 10^{21} <
n_{e} < 10^{24} {\rm cm}^{-3}$, $10^{4} {\rm K} < T < 10^{6} { \rm
K}$. The results are compared to some theoretical data.

\end{abstract}
PACS{: 51.10.+y, 51.70.+f, 52.25.Fi, 52.27.Gr, 52.80.Pi}

\section{Introduction}
The determination of the (internal) dynamic conductivity (i.e., the
response to the homogeneous high-frequency Maxwellian electrical
field $\vec{E}(t)= \vec{E_{0}}{\rm exp}(-i\omega t))$ of dense
plasmas has been a subject of substantial investigation for a long
time. One of the reasons is that on the basis of this quantity all
other plasma dynamic characteristics can be found \cite{kobzev_5}.
There are two basic approaches to these studies: the generalized
Drude-Lorentz model, see \cite{hr1}, the review \cite{hr2} and
references therein, and the method of moments \cite{Tkach}.
Additionally, we have been working on the direct extension of the
modified random-phase approximation for the calculation of the
static conductivity $\sigma _{0}$, \cite{jphysd2004}.

Previously, in \cite{jphysd2001}, we applied the latter approach in
the range of slightly and moderately non-ideal plasmas with the
number density of electrons $n_{e}$ and temperature varying within
the following limits: $ 10^{17} < n_{e} < 10^{19}{ \rm cm}^{-3}$,
$10^{3}{ \rm K} < T < 10^{4}{ \rm K}$ and examined the range of
frequencies that covered the microwave and far-infrared region. In
\cite{jphysd2004,jphysA1} we extended the range of
frequencies up to the ultraviolet radiation and covered the area of
very high values of the electron density: $10^{21} <n_{e}<10^{23}{
\rm cm}^{-3}$ with $10^{3}{ \rm K} < T < 10^{5}{ \rm K}$.

The classical Drude-Lorentz formula for the plasma optical conductivity,
\begin{equation}
\sigma _{DL}(\omega )=\frac{\sigma _{0}}{1-i\omega \tau },\qquad
\tau =\frac{ 4\pi \sigma _{0}}{\omega _{p}^{2}},
\label{eq::sigma_din4}
\end{equation}
where $\sigma_{0}$ is static conductivity, and $\omega
_{p}=\sqrt{4\pi n_{e}e^{2}/m}$ being the plasma frequency, predicts
a monotonic decrease of its real part when the frequency $\omega
\rightarrow \infty $, and it is not clear whether this property is
maintained in real dense plasmas.

In the present work we study the question of monotonicity of the real part
of the dynamic conductivity in even wider ranges of variation of the plasma
parameters.

\section{The model}

We consider the (internal) dynamic conductivity of hydrogen plasmas
in a volume $V$ containing $N_{e}=n_{e}V$ electrons and the same
number of ions.

As a starting point for the computations we use the exact relation
for the optical conductivity of Coulomb systems stemming from the
theory of moments \cite{tvt}
\begin{equation}
\sigma (\omega )=\frac{i\omega _{p}^{2}}{4\pi }\frac{\omega
+q(\omega )}{ \omega ^{2}-\Omega ^{2}+\omega q(\omega )},
\label{eq::sigma_din1}
\end{equation}
where $q(\omega )$ is the boundary value of some analytic
(Nevanlinna)
function $q(z)$, which admits the representation%
\begin{equation}
q(z)=ih+2z\int_{0}^{\infty }\frac{du(\omega )}{\omega ^{2}-z^{2}},
\label{eq::q}
\end{equation}
with $h\geq 0$ and a non-decreasing bounded function $u(\omega )$
such that
\begin{equation*}
\int \limits_{-\infty }^{\infty }\frac{du(\omega )}{1+\omega
^{2}}<\infty .
\end{equation*}
Independently of the choice of $ q\left( z\right) $, the optical
given by the expression (\ref{eq::sigma_din1}), has the following
exact asymptotic expansion \cite {tvt}
\begin{equation}
\sigma (\omega \rightarrow \infty )\simeq \frac{i\omega
_{p}^{2}}{4\pi \omega }+\frac{i\omega _{p}^{2}\Omega ^{2}}{4\pi
\omega ^{3}}+o(\frac{1}{ \omega ^{3}}).  \label{eq::sigma_din2}
\end{equation}
The estimates for the characteristic frequency $\Omega $
 \cite{tvt} are provided in the next section. The parameter function
$q(z)$ possesses no phenomenological meaning, but we can observe
that the condition $\lim \limits_{\omega \rightarrow 0}q(\omega
)=ih=4\pi i\sigma _{0}\left(\Omega/\omega_{p}\right) ^{2}$ is
equivalent to the definition $\lim \limits_{\omega \rightarrow
0}\sigma (\omega )=\sigma _{0}$. Hence, the simplest formula
providing an interpolation between the exact asymptotic expansion
(\ref {eq::sigma_din2}) and the static conductivity has the
following form:
\begin{equation}
\sigma (\omega )=\frac{i\omega _{p}^{2}}{4\pi }\frac{\omega +i\tau
\Omega ^{2}}{\omega ^{2}-\Omega ^{2}+i\omega \tau \Omega ^{2}}.
\label{eq::sigma_din3}
\end{equation}

We have previously calculated the plasma static conductivity in a
wide range of plasma thermodynamic parameters, see \cite{jphysd2001,jphysd2004}. 
We used this data and also carried out
additional computations of $\sigma _{0}$ using the same
self-consistent field method \cite{tvt1} (for recent results
obtained using this approach see \cite{juan}) to find the values of
the transport relaxation time $\tau $ in an extended realm of the
$n_{e}-T$ plane. Certainly, to evaluate the static conductivity one
can employ alternative theoretical approaches like that of
\cite{redmer}.

Notice that (\ref{eq::sigma_din3}) turns into the classical
Drude-Lorentz formula when $\Omega ^{2}\rightarrow \infty $, i.e,
when the asymptotic expansion (\ref{eq::sigma_din2}) reduces to that
satisfied by the Drude-Lorentz dynamic conductivity, $\sigma
_{DL}(\omega \rightarrow \infty )\simeq i\omega _{p}^{2}/4\pi \omega
+o(\omega ^{-1}).$

\section{The parameter $\Omega ^{2}$}

To estimate the dimensionless parameter
\[
H=\Omega ^{2}/\omega _{p}^{2}=h_{ei}\left( 0\right) /3=\left( 2\pi
^{2}n_{e}\right) ^{-1}\int_{0}^{\infty }k^{2}S_{ei}\left( k\right)
dk,
\]
at least in strongly coupled hydrogen plasmas, one can use the
interpolation procedure suggested in \cite{tvt}: approximate the
static electron-ion structure factor $S_{ei}\left( k\right) $ at a
zero Matsubara frequency \cite{AGD}
\begin{equation}
S_{ei}\left( k\right) =P_{e}\left( k\right) P_{i}\left( k\right)
/\left[ k^{2}\lambda ^{2}+P_{e}\left( k\right) +P_{i}\left( k\right)
\right] , \label{m5}
\end{equation}
but, to go beyond the RPA, put the ion and electron dimensionless
polarization operators (simple loops) as%
\begin{equation}
P_{i}\left( k\right) =\Pi _{i}(k)/n_{e}\beta =1,\quad P_{e}\left(
k\right) =\Pi _{e}(k)/n_{e}\beta =\gamma ^{4}\lambda ^{2}/\left(
k^{2}+\gamma ^{4}\lambda ^{2}\right) ;  \label{pii}
\end{equation}%
this interpolation being constructed to satisfy both the long- and
short-wavelength limiting conditions \cite{BBT,tvt}:
\begin{equation}
P_{e}\left( k=0\right) =1,\quad P_{e}\left( k\longrightarrow \infty
\right) \simeq \gamma ^{4}\lambda ^{2}/k^{2}  \label{pieq0}
\end{equation}
with
\begin{equation}
\gamma ^{4}=16\pi n_{e}e^{2}m/\hbar ^{2},\quad \lambda ^{-2}=4\pi
e^{2}n_{e}\beta .  \label{gamma4}
\end{equation}
Then by simple integration one gets \cite{mj}:
\begin{equation}
H=\left( 4r_{s}/3\right) \sqrt{\Gamma /\left( 3\Gamma
^{2}+4r_{s}+4\Gamma \sqrt{6r_{s}}\right) }.  \label{Hnew}
\end{equation}
Observe that in weakly coupled plasmas with $\Gamma \longrightarrow
0$,
\begin{equation}
H\simeq \left( 2/3\right) \sqrt{r_{s}\Gamma }\sim \sqrt{\beta }.
\label{hei0g0}
\end{equation} \label{b}


\begin{table}
  \centering
  \caption{The dimensionless static conductivity $\sigma_{0}/\omega _{p}$
  as a function of electron density and temperature.}\label{tab::sig0_RPA}
  \begin{tabular}{|c|c|c|c|}
    \hline
    \hline
    $n_{e}\cdot10^{-22}({ \rm cm}^{-3})$ & $T=2\cdot10^{4}{ \rm K}$ & $T=3\cdot10^{4}{ \rm K}$ & $T=5\cdot10^{4}{ \rm K}$ \\
    \hline
    \hline
    $1$ & $0.23$ & $0.26$ & $0.32$ \\
    \hline
    $10$ & $0.36$ & $0.38$ & $0.41$ \\
    \hline
    $100$ & $0.78$ & $0.80$ & $0.81$ \\
    \hline
    \hline
  \end{tabular}
\end{table}
\begin{figure}[h!]
\centering
\begin{minipage}[t]{0.49\textwidth}
      \centering
      \includegraphics[width=\textwidth, height=0.75\textwidth]{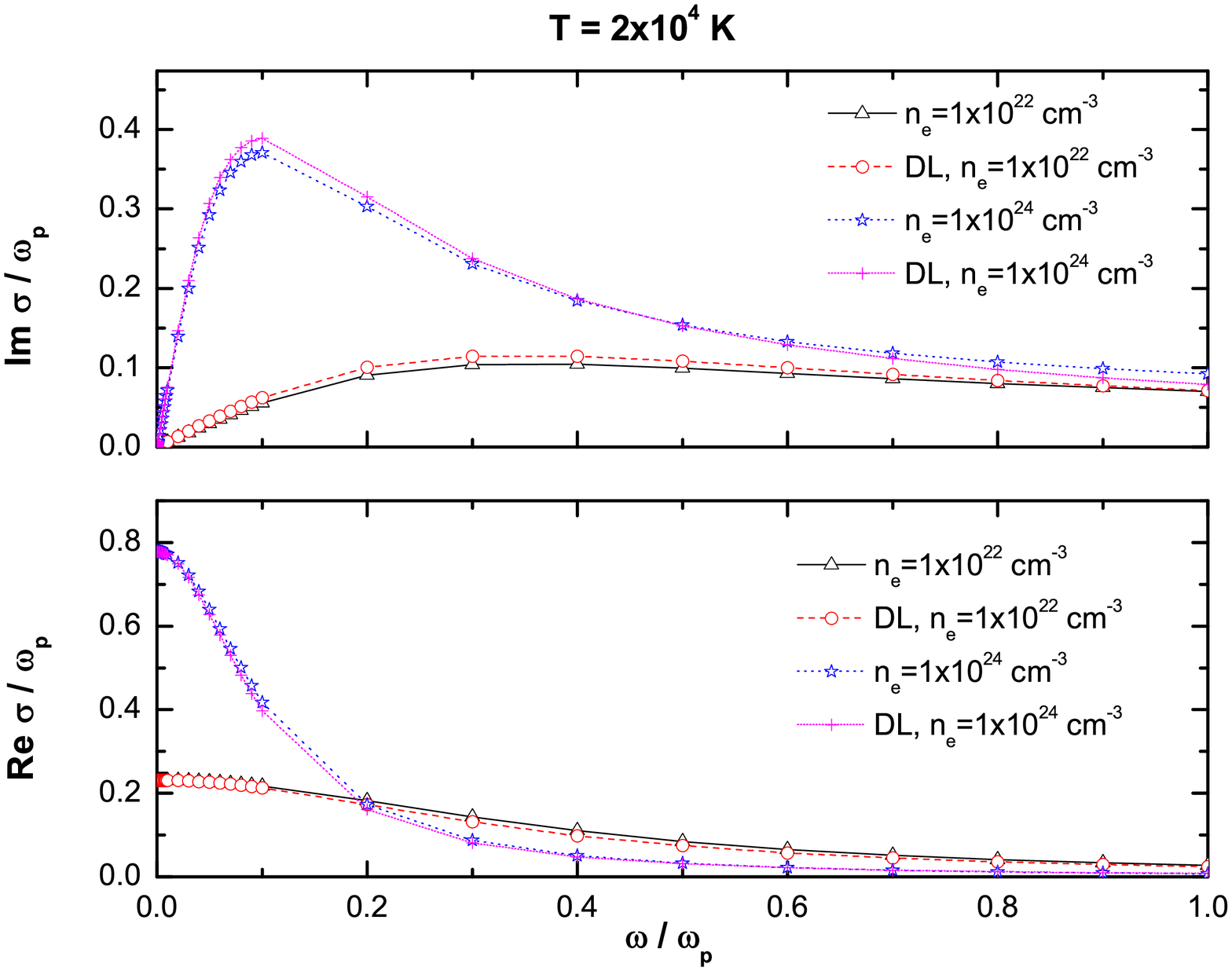}
    \end{minipage}
\hfill
\begin{minipage}[t]{0.49\textwidth}
      \centering
      \includegraphics[width=\textwidth, height=0.75\textwidth]{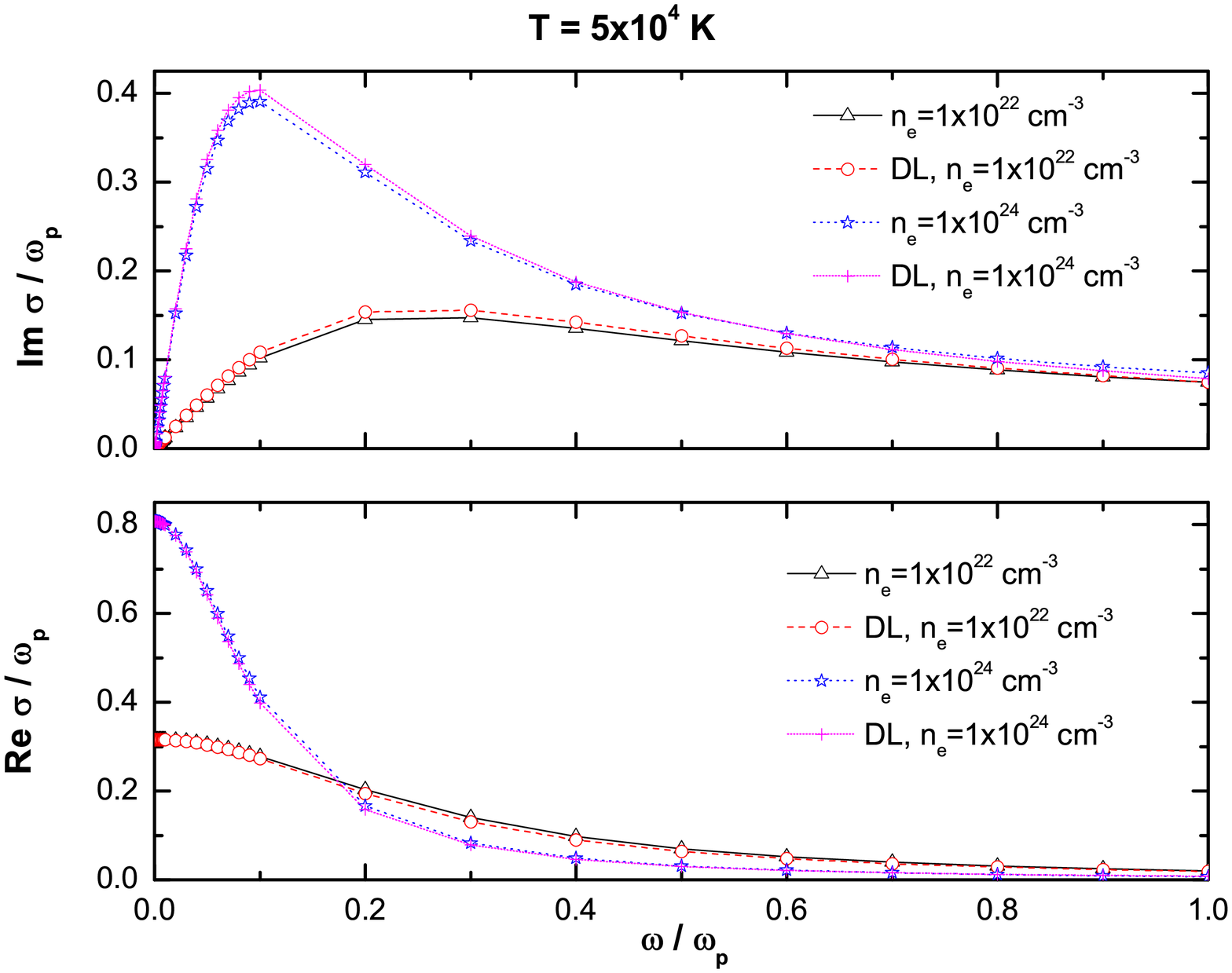}
    \end{minipage}
\caption{Dynamic conductivity of dense plasmas according to
(\ref{eq::sigma_din3}) compared to the Drude model
(\ref{eq::sigma_din4}) prediction. The static conductivity is
provided in table \ref{tab::sig0_RPA}.} \label{fig::30000}
\end{figure}
\begin{figure}
  \begin{center}
    \begin{tabular}{cc}
      \resizebox{60mm}{!}{\includegraphics{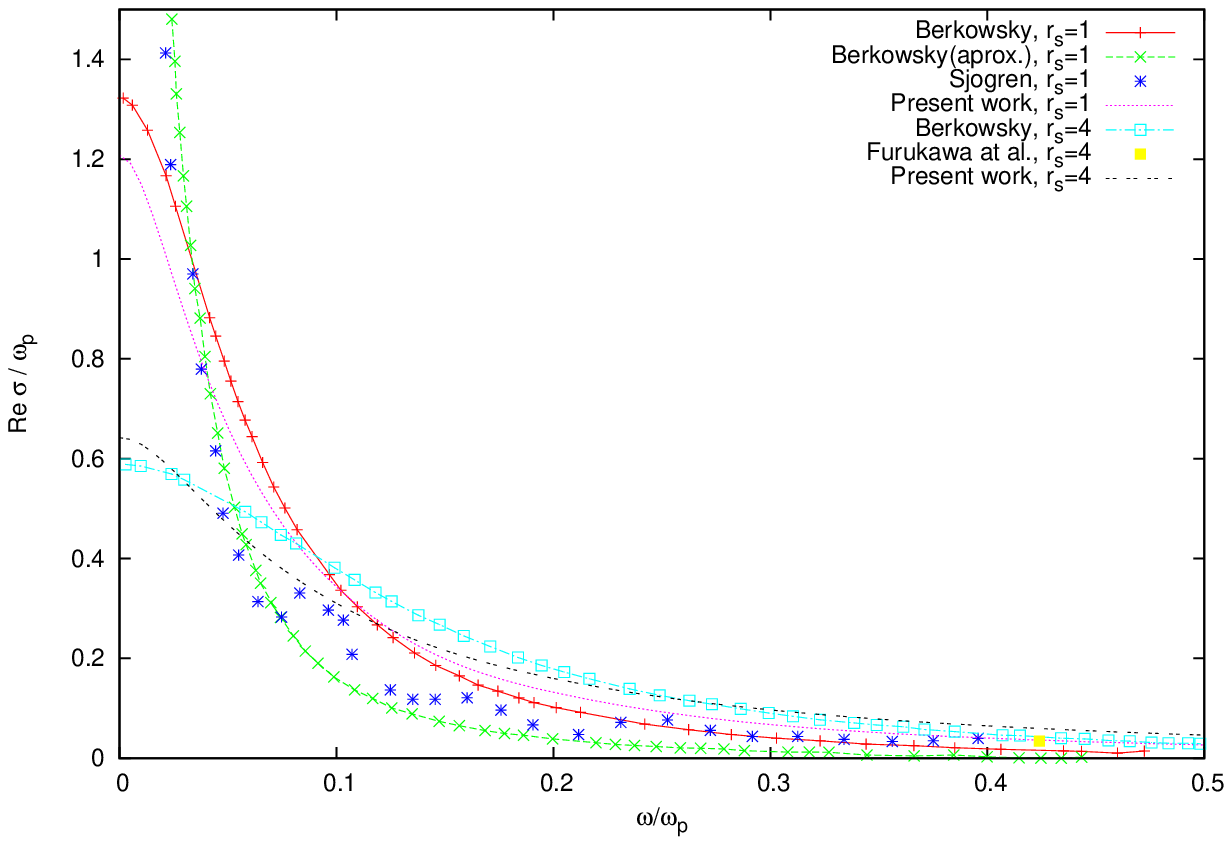}} &
      \resizebox{60mm}{!}{\includegraphics{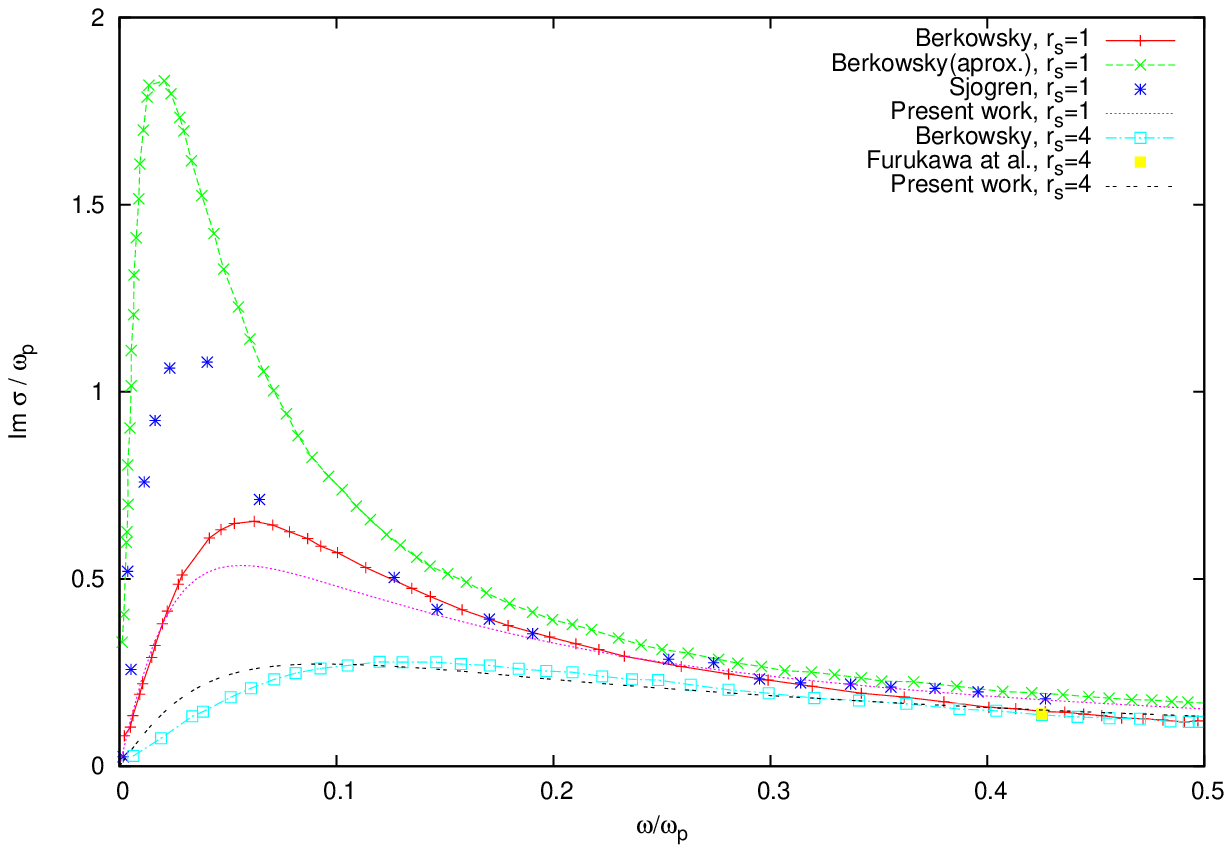}} \\
    \end{tabular}
    \caption{Data calculated from (\ref{eq::sigma_din3}) for $\Gamma =
0.5, r_{s} = 1$ and $r_{s} = 4$, as a function of ($\omega /
\omega_{p}$) frequency ratio, together with results of other authors
\protect \cite{berkovski}.}
    \label{fig::berk1}
  \end{center}
\end{figure}

\section{Results and discussion}

The calculations of the real and imaginary parts of $\sigma(\omega)$
in (\ref{eq::sigma_din3}) were carried out in the domain
$10^{21}\leq n_{e}\leq 10^{24}{ \rm cm}^{-3}$ and $10^{4}{ \rm
K}\leq T\leq 10^{6} { \rm K}$.

In the figure \ref{fig::30000} we compare some of these results to
those corresponding to the Drude-Lorentz model
(\ref{eq::sigma_din4}). For the reference we provide also the data
for the dimensionless static conductivity $\sigma_{0}/\omega _{p}$,
see table \ref{tab::sig0_RPA}.

We observe that within the present model no qualitative difference
exists between our results and those of the Drude-Lorentz model
(\ref {eq::sigma_din4}). Quantitative difference decreases as
$\Gamma \rightarrow 0.$ It is evident that whenever
$\xi=\left(1-\tau ^{2}\Omega ^{2}/2\right)>0$, the real part of
(\ref{eq::sigma_din3}) acquires an additional maxima at
$\omega_{m}=\pm \Omega \sqrt{\xi}$, but for our data the values of
$\xi $ are always negative.

Additionally, we successfully compare the data on
$\sigma(\omega;n_{e},T)$ determined in this paper to the data from
\cite{berkovski} for $\Gamma =0.5$, $r_{s}=1$ and $r_{s}=4$. The
corresponding curves are shown in Fig. \ref{fig::berk1}.

Detailed comparison of our results to those of other approaches, in
particular, those described in \cite{hr2}, is due. We conclude that
our results can be used for the investigation of dynamic and static
properties of strongly coupled plasmas.

\section{Acknowledgments}

The presented work is performed within the Project 141033 financed
by the Ministry of Science of Republic Serbia and is supported by
the Spanish Ministerio de Ciencia e Innovaci\'{o}n (Project No.
ENE2007-67406-C02-02/FTN) and the INTAS (GSI-INTAS Project
06-1000012-8707).

\section{References}

\bibliographystyle{unsrt}

\end{document}